# SRPS: Secure Routing Protocol for Static Sensor Networks

Hamoinba Vebarin and Samourqi Difrawi
International Awareness Institute

Email: {hvebarin, sdifrawi}@iwi.org.jp

## Abstract

In sensor networks, nodes cooperatively work to collect data and forward it to the final destination. Many protocols have been proposed in the literature to provide routing and secure routing for ad hoc and sensor networks, but these protocols either very expensive to be used in very resource-limited environments such as sensor networks, or suffer from the lack of one or more security guarantees and vulnerable to attacks such as wormhole, Sinkhole, Sybil, blackhole, selective forwarding, rushing, and fabricating attacks. In this paper we propose a *secure lightweight* routing protocol called SRPS. SRPS uses symmetric cryptographic entities within the capabilities of the sensors, supports intermediate node authentication of the routing information in addition to end-to-end authentication, provides secure multiple disjoint paths, and thwarts all the known attacks against routing infrastructure against Byzantine cooperative attack model. We analyze the security guarantees of SRPS and use Ns-2 simulations to show the effectiveness of SRPS in counter-measuring known attacks against the routing infrastructure. Overhead cost analysis is conducted to prove the lightweight-ness of SRPS.

**Keywords**: sensor network security, secure routing, symmetric cryptography, multiple disjoint paths, neighbor watch, active and passive attacks.

## 1  Introduction

The open nature of the communication media used in wireless ad-hoc and sensor networks, the lack of infrastructure, the fast deployment polices, and the hostile environments where they usually deployed in, make these networks vulnerable to a wide range of attacks. Of the many areas vulnerable to attacks in sensor and ad hoc networks are the routing protocols. Attacks on routing can be external as well as internal, and this means that there is a need to come up with schemes to safeguard the routing process. Authentication of the routing information by the nodes involved in the route is necessary to prevent this information from being fabricated or modified by compromised nodes. Internal attacks on the routing infrastructure include Wormhole attack, Rushing attack, Sybil attack, Selective forwarding, Sinkhole attack, spoofing, and malicious forge or change of the routing information.

Due to the constrained resources available to sensor networks, the application of known countermeasures used in wired networks will not be applicable. Also, while wireless sensor networks share similarities with ad-hoc wireless networks, the dominant communication method on both is multi-hop; there are several important distinctions. Firstly, ad-hoc networks usually support routing between any pair of nodes in the network, whereas sensor networks has a narrower communication pattern in which the bulk of the traffic is between a centralized point (called sink or base station) and each sensor in the network, and little traffic goes between sensors, usually within the same geographic vicinity for coordination and data aggregation. Secondly, the resources in sensor networks (energy, CPU, memory, storage, and bandwidth) are far more limited than those of ad-hoc networks. Finally, due to redundancy in sensor networks, many neighboring sensors observe the same or correlated environmental events. If every one of these events is going to be sent to the base station independently, then precious resources will be wasted. Thus, there must be trust relationships among sensors beyond those typically found in ad-hoc networks, to cooperate in data aggregation and duplicate elimination to optimize resource usage.

*Contribution:* in this paper we present a secure routing protocol for static sensor networks, called *SRPS* that:

1. Thwarts the internal Byzantine attacks launched by compromised nodes such as wormhole attack, Sybil attack, and blackhole attack.
2. Presents a novel way to provide per-hop authentication of the routing information in addition to the end-to-end authentication.
3. Is lightweight and only uses symmetric key cryptography tools to the extent of sensors capabilities.
4. Does not require any special hardware (such as directional antennas or GPS).
5. Does not require any time synchronization among the nodes in the network (neither tight nor loose).
6. Provides an idea to link short commitment sequences without the need to provide a new commitment key when the current commitment sequence is exhausted, this eliminates the need for large commitment sequences.
7. Supports secure multiple-disjoint-path discovery between the two end points of communication.

Trivial Denial of Service attacks based on interception and non-cooperation exist in all ad hoc routing protocols but they are not achieved through subversion of the routing protocols, so they will not be considered.

The rest of the paper is organized as follows. Section 2 presents the related work in the field of secure routing protocols in wireless ad-hoc and sensor networks. Section 3 describes the SRPS protocol.

Section 4 presents security analysis of SRPS. Section 5 presents coverage and cost analysis of SRPS. Section 6 presents simulation results. Finally, Section 7 concludes the paper.

## 2  Related Work

Many ad-hoc network security mechanisms for authentication and secure routing protocols have been proposed in the literature. Some of them are based on local monitoring ([1] - [6]), while others are based on the public/symmetric key cryptography ([7]-[13]). Those protocols either suffer from weak security measures or have a very expensive overhead which sensor networks can not afford.

Papadimitratos and Haas [13] present the SRP protocol that is secure against non-colluding adversaries by disabling route caching and providing end-to-end authentication using an HMAC primitive. SEAD [10] uses one-way hash chains to provide authentication for DSDV [27]. Ariadne [11] uses an authenticated broadcast technique [44] to achieve similar security goals on DSR [29]. Marti *et. al.* [34] examine techniques to minimize the effect of misbehaving nodes through node snooping and reporting, but it is vulnerable to blackmail attacks. ARRIVE [31] proposes probabilistic multi-path routing instead of single path algorithm to enhance the robustness of routing. These secure routing protocols are still vulnerable to wormhole [48] attacks that can be conducted without having access to any cryptographic keys.

Zhu *et. al.* [38] present LHAP, an authentication protocol for ad hoc networks. LHAP is based on a hop-by-hop authentication for verifying the authenticity of all the packets transmitted in the network and on one-way key chain and TESLA for packet authentication. LHAP uses also asymmetric key cryptography to bootstrap trust in the network. All these tools are highly expensive for sensors making LHAP infeasible in sensor networks.

It is usually infeasible to apply the above-mentioned protocols to sensor networks. The public key cryptography is far beyond the capabilities of sensor nodes. And the symmetric key protocols proposed are too expensive in terms of node state and communication overhead. Many sensor network routing protocols have been proposed ([14], [15], [16], [17], [18],[19], [20]) and many secure applications have been proposed, but many of them are even more susceptible to attacks against their routing infrastructure. These attacks fall into one or more of the following categories: spoofed, altered, or replayed routing information, selective forwarding, sinkhole attacks, Sybil attacks, wormholes, HELLO flood attacks, and acknowledgement spoofing.

Karlof and Wagner [33] analyze the vulnerability of various routing protocols to one or more of the above mentioned attacks and provide general framework for countermeasures of these attacks. However, they did not propose a secure solution for routing but leave it as an open problem to design a sensor network routing protocol that satisfies the security goals they propose. They show that TinyOS beaconing suffers from bogus routing information, selective forwarding, sinkholes, Sybil, wormholes, and HELLO flood attacks. Directional diffusion [14] and its multi-path variant [17] suffer from bogus routing information, selective forwarding, sinkholes, Sybil, wormholes, and HELLO flood attacks. Geographical routing protocols, GPSR [16] and GEAR [18], suffer from bogus routing information, selective forwarding, and Sybil attacks. Minimum cost forwarding [19] suffers from bogus routing information, selective forwarding, sinkholes, wormholes, and HELLO flood attacks. Clustering based protocols, LEACH [20], TEEN [40], PEGASIS [41], suffer from selective forwarding and HELLO flood attacks. Rumor routing [21] suffers from bogus routing information, selective forwarding, sinkholes, Sybil, and wormhole attacks. Wireless reprogramming protocols ([26][25]) suffer from high overhead. Other domain protocols also ( [42] [43]) also has high overhead.

Hu *et. al*. [37] describe the rushing attack in wireless ad hoc network routing protocols and propose a countermeasure through dynamic secure neighbor detection, secure route delegation, and randomized route request forwarding. This protocol depends on a very strict time propagation delay measurement to detect neighbors. Neighbor detection is done on the fly and for every packet exchanged which may affect the efficiency and the ability of using it in highly constrained sensor networks. Hu and Evans [36] propose a solution to the wormhole attacks using specialized hardware; directional antennas

TESLA [28] and µTSLA [44], use an authentication technique that uses periodic and delayed key disclosure. Delayed authentication (as in TESLA, and µTESLA) is not appropriate since a packet would be delayed at each node in the path from the source to the destination. Moreover, since each node has to buffer the traffic packets it has received until they are authenticated, delayed authentication will lead to high storage requirement at every node. However, *SRPS* uses an authentication technique that does not need a large storage by a voiding the pre-computation of a hash sequence of keys and storing them in advance for the whole life time of the sensor. Also *SRPS* authentication doesn't require any kind of time synchronization because the authentication is limited to the neighborhood and all the neighbors receive the data at the same time, so there is no chance for any of them to forge messages and broadcast them on behalf of the real source.

Some routing protocols (e.g. [27], [29]) use non-repeating increasing counters for the route request and route reply packet identifier while others use random numbers (e.g. [13], [37]). However, these ways for the packet identifier add vulnerabilities to the routing protocol. If an increasing sequence number is used, an attacker can track the sequence number of the route requests from any source, say S, to any destination, say D, and then launches a DoS attack by flooding the network with a higher number. This will prevent S and D from discovering any new routes between them, since all the nodes in the network will think that this is an old request and just drop it. On the other hand, if a random number is used, an attacker can easily replay old attacks since the destination can't distinguish between an old valid request and a new one using the random number alone.

Both of these choices fail to countermeasure the malicious inclusion of compromised nodes in an already established route [39]. The success of this attack facilitates the success of other attacks such as the blackhole attacks and the selective forwarding attacks. A compromised node can include itself in an already established route by sending a valid route request (correct sequence number or random number) to a node already in the route. Figure 1 shows an example of how can a malicious node includes itself in an already established route in a protocol like DSDV [28]. The malicious node, M, sends to Y a route request with a source sequence number greater than the current one. In response to that, Y changes the route to A to point to M instead of X. Then M sends to X a route reply packet with a destination sequence number greater than the current one. X changes its route to B to point to M instead of Y. Thus M succeeds in including itself in the route easily.

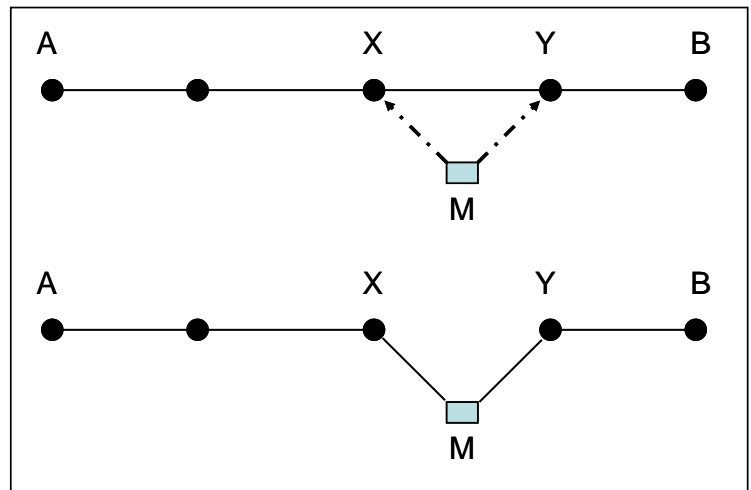

**Figure 1: (A) An example of a malicious node broadcasting fabricated route request. (B) The route after the malicious node succeeds**

To countermeasure this attack, there must exist a method by which an intermediate node can verify the authenticity of the route request and the route reply. ARAN [39] proposes a solution using public key cryptography with centralized trusted authority and signatures. The route initiator signs the route request packet and every intermediate node verifies the signature of the initiator verifies the signature of the previous hop, replaces the signature of the previous hop with its own signature. Ariadne [11] suggests three methods to provide authenticity to the intermediate nodes; using pair-wise symmetric keys in which the route request/reply must be individually authenticated to each intermediate node and to destination/source, using TESLA protocol which provides authenticate public broadcasting, and uses digital signatures. These solutions for intermediate node authentication are very expensive in terms of the computational and bandwidth overhead which makes them impractical for a highly limited resources sensor networks.

Neighbor watch has been used as an enhancement of the watch dog approach [34], which was used to negate the effect on throughput of misbehaving nodes that agree to forward packets but do not. But watch dog suffers from problems such as ambiguous collisions and receiver collisions. A malicious watcher could blackmail a good node by claiming that it did not forward a packet while it is really did, or it may not receive the message due to collision. Another problem could occur when the watcher is not able to detect whether the receiver forwards the packet or not if the watcher got jammed by the time when the receiver forwards. Neighbor watch mitigates these problems by not limiting the watch to the packets' sender but asking other neighbor nodes to work as watchers.

## 3  Description of SRPS

*Overview:* Immediately after the deployment of the sensors in the network, the initial phase (Section 3.2) starts, this provides each node with a list of neighbors, a commitment key for each neighbor, and the neighbors of each neighbors (first and second hop neighbors). This neighbor list is built using neighbor detection protocols (Section 3.1). Anode that needs to communication with another node checks its routing table; if a route does not exist, it enters a route discovery phase (Section 3.3).  In the route discovery phase, the route request initiator floods the network with a route request which propagates to the destination. The destination unicasts a route reply back to the initiator which stores the route in its routing table, leaves the route discovery phase, and starts using the established route. Each intermediate node in path of the route reply verifies the authenticity of the route reply (Section 3.4), updates its routing table and forwards the packet to the next hop. Each node that forwards the route request or the route reply proves its identity to its neighbors through MAC authentication using its commitment key (Section 3.2). Each node that can overhear the route reply monitors the behavior of the nodes involved in forwarding the route reply for suspicious actions, such as fabrication, change, or dropping (Section 3.5).

Some applications (e.g. secure data transmission) require multiple disjoint paths to exist between the source and the destination, Section 3.6 explains how *SRPS* provides disjoint multiple paths. If an established route breaks, the first intermediate node that notices the break sends a route error packet back to source, in a process called route maintenance (Section 3.7).

*Assumptions:* We assume that the links are bi-directional; which means that if a node A can hear node B then node B can hear node A. Also we assume the existence of an underlying pair-wise key management protocol ([44], [45], [46]). Any two nodes willing to establish a route between them are assumed to have a joint shared secret key, distributed using the underlying key-management protocol. Finally we assume that there are no malicious nodes during the setup phase of the network. We assume a none-mobile dynamic topology sensor network, i.e. the nodes do not move but the roles they play in the network (e.g. sensing role, cluster head, control node, data aggregator …) are changed.

*Attack Model:* A full cooperative Byzantine attack model is considered in which compromised nodes can collude to do whatever they can to subvert the functionality of the routing protocol. A node only trusts itself and the main base station.

## 3.1 Neighbor Detection

Neighborhood detection can be achieved through multiple ways:

Centralized information from the base station that knows the topology of the network and keeps updating that topology in mobile networks. This can be amortized with protocols that introduce control nodes or cluster heads for scalability and efficiency.

Naïve detection by broadcasting a HELLO packet. But this method is vulnerable to wormhole attacks where powerful compromised nodes can fool other nodes to believe that they are neighbors even though they may be multiple hops away. Wormhole prevention techniques can be used to prevent this attack [36].

Directional-antenna-based neighborhood detection which is used as well to prevent Wormhole attacks [36].

Propagation delay neighborhood detection techniques where packet delay of certain control packets is used to measure the distance to a neighbor [37].

Under the assumptions that we consider, the neighborhood discovery is done only once at the setup phase and is guaranteed to be secure. So we will use the second method to establish the neighbor list of each node, which will be simple and safe since it is built at the time when no malicious nodes are assumed to exist. *SRPS* however, could be easily extended to dynamic and mobile networks if it is augmented with one of the other three neighbor detection techniques mentioned above (0, 0, and 0).

After the node gets the list of its direct neighbor, it exchanges this list with each neighbor so that each node stores its direct neighbors and the neighbors of each neighbor.

## 3.2 Initial Setup

As soon as the sensor nodes are spread in the field, each node starts sending a HELLO message which is replied to by all the nodes that hear it. Both the HELLO packet and its reply are small packets that contain only the address of the packet initiator. For each reply to the HELLO packet, the initiator of the HELLO packet adds the source of the reply to its neighbor list. This processes is performed only once in the whole lifetime of the sensor network, so the associated overhead is affordable. Upon completing, each sensor node *i* have a list of its neighbors, which we denote by $R_i$.

In addition to the knowledge of the neighbors, each node needs a mechanism to authenticate each one of them. To achieve this, each sensor node, say S, distributes a commitment key to every node in $R_S$. This can be done in two ways. In the first, the commitment key is exchanged during the setup phase using the HELLO messages. Each HELLO massage is augmented with a commitment key ($K_{commit(S)}$) from the sender (S) to each neighbor who stores the commitment before replying to the HELLO. In the second, the commitment key is exchanged using the underlying key management protocol. As a result, each node, say S, ends up having the commitments of its neighbors as well as its own commitment sequence seed ($K_{seed(S)}$).

The commitment key is derived from the commitment seed as $K_{Commit(S)} = F^{(t)}(K_{seed(S)})$, where F is a one-way collision resistant function and *t* is the length of the commitment string. The value of *t* depends on the amount of available memory. It can be as small as two and as big as desired. However, the longer the sequence the lower the communication overhead incurred by commitment renewal (Section 3.4.3) and the larger the amount of memory required. So the tradeoff is between the communications overhead versus the storage requirement. To overcome the storage requirement, the commitment key can be derived on demand by applying the one-way function to the commitment seed ($K_{seed(A)}$) multiple times as needed. For example, the j$^{th}$ authentication key = $F^{(t-j)}$ (). So the tradeoff becomes between the memory requirement and the computational cost.

When a node B wants to broadcast an authentic packet to its neighbors, it first generates the next authentication key as $K_{auth(B)}$ = F(last known authentication key), and uses it to generate a MAC over the packet. The first authentication key = F ($K_{seed(S)}$). B then broadcasts the packet to its neighbors. After all the neighbors got the packet, B broadcasts its current authentication key, $K_{auth(B)}$, that is used to authenticate the previously sent packet. When B releases $K_{auth(B)}$, each neighbor, say A, verifies the validity of the key by running the hash function over it and comparing the result with the stored commitment for B. If the key is valid, A stores it as the new commitment for B and uses it to verify the authenticity of the packet it received from B.

An alternative neighbor authentication could be achieved if we assume that the hardware addresses can't be controlled by the attacker. If we set the ID of each node to be the hash value of the hardware address ($ID_X$ = F(Hardware_address_of X)), then a neighbor node can find the ID of the source of the packet by calculating it from the hardware address of the source. This also has the benefit of smaller packet overhead since the ID of the source of the packet need not to be included in the packet header.

## 3.3 Route Discovery

When a node, say S, needs to discover a route to a destination, say D, it generates a route discovery packet (RDP) that contains: a flag to indicate that it is a route request packet (REQ), the sender ID ($ID_S$), the destination ID ($ID_D$), a unique sequence number (SN), and a sequence number verification (SNV), see Section 3.4.1 for the generation of SN and SNV. S then calculates a MAC value over the packet using the shared key between S and D ($K_{SD}$). S then calculates the neighborhood MAC using its current neighborhood authentication key $K_{auth(S)}$. Finally, S broadcasts the packet to its neighbors.

- S → generates: $RDP = REQ \| ID_S \| ID_D \| SN \| SNV$
- S → calculates: $MAC_{K_{SD}}(RDP)$
- S → calculates: $MAC_{K_{auth(S)}}(RDP \| MAC_{K_{SD}}(RDP))$
- S → Broadcast: $RDP \| MAC_{K_{SD}}(RDP) \| MAC_{K_{auth(S)}}(RDP \| MAC_{K_{SD}}(RDP))$

Each neighbor of S receives the broadcast and stores it for verification. S then broadcasts $K_{auth(S)}$ to its neighbors. Each neighbor verifies the authenticity of $K_{auth(S)}$ by calculating $F(K_{auth(S)})$ and comparing it with the stored commitment for S. If $K_{auth(S)}$ passes the check, it is used to verify the integrity of the previously received packet by calculating the MAC value $MAC_{K_{auth(S)}}(RDP \| MAC_{K_{SD}}(RDP))$ and comparing it with the MAC value associated with the packet, and $K_{auth(S)}$ replaces the old commitment key. A node B in $R_S$ waits for a random time, $T_r$, selected from $[T_{min}, T_{max}]$, during this time it collects every route broadcast it could hear from a different neighbor. When $T_r$ runs out or when a certain number of requests, $N_r$, is collected, whichever occurs first, B selects at random one of the requests it has in its buffer and suppresses the rest. Assume without loss of generality that B selects the one it heard from S. B then removes the $MAC_{K_{auth(S)}}$ part of the message, appends its ID ($ID_B$) and the ID of the node from which it receives the packet ($ID_S$ for B since it receives it directly from the source), appends its own authentication, and re-broadcasts the packet to its neighbors. The process continues the same way until the packet reaches the destination R.

- B → stores: $RDP \| MAC_{K_{SD}}(RDP) \| MAC_{K_{auth(S)}}(RDP \| MAC_{K_{SD}}(RDP))$
- S → broadcasts: $K_{auth(S)}$
- B → keeps in its buffer: $ID_S$, $ID_D$, SN, SNV, the ID of the node from which it receives the packet, and the ID of the second-hop previous sender of the packet.
- B → waits for $T_r$ during which it repeats the previous step for every received broadcast.

B → selects at random one of the packets in its buffer

B → broadcasts: RDP || $MAC_{K_{SD}}$(RDP) || $ID_B$ || $ID_S$ || $MAC_{K_{auth(B)}}$(RDP|| $MAC_{K_{SD}}$(RDP) || $ID_B$ || $ID_S$)

When D receives the packet, it verifies the authenticity of the source using the shared key $K_{SD}$. Then D generates a route reply packet RRP that contains: a flag to indicate that it is a route reply packet (REP), the sender ID ($ID_S$), the destination ID ($ID_D$), a SN and a SNV. D then calculates a MAC value over the packet using the shared key between S and R ($K_{SR}$). Finally D calculates the neighborhood MAC using its current neighbor commitment key ($K_{auth(D)}$). And if A is the node from which D receives the corresponding route request which received it from C, D sends the route reply packet back to A to be forwarded to C.

A → broadcasts: RDP || $MAC_{K_{SD}}$(RDP) || $ID_A$ || $ID_C$ || $MAC_{K_{auth(A)}}$(RDP|| $MAC_{K_{SD}}$(RDP) || $ID_A$ || $ID_C$)

D → generates: RRP = REP || $ID_S$ || $ID_R$ || SN || SNV

D → calculates: $MAC_{K_{SD}}$(RRP)

D → calculates: $MAC_{K_{auth(D)}}$(RRP|| $MAC_{K_{SD}}$(RRP))

D → A: RRP || $MAC_{K_{SD}}$(RRP) || $MAC_{K_{auth(D)}}$(RRP|| $MAC_{K_{SD}}$(RRP))

Node A stores the route reply packet it received from D. D then broadcasts the commitment key $K_{auth(D)}$. Each node in the neighborhood of D, $D_R$, verifies $K_{auth(D)}$ and updates the commitment of D they already have. A then uses $K_{auth(D)}$ to verify the integrity of the route reply by calculating the MAC and comparing it with the MAC associated with the packet. A verifies the authenticity of the route reply and the route request, Section 3.4.1. If the request-reply pair passes the verification, A updates its routing table to S and to D. A then removes the $MAC_{Kauth(D)}$ part of the message, appends its ID, appends the ID of the second-way hop, C, that the reply should be forward to, associates its own authentication, and sends the packet back to C. The process continues the same way until the reply reaches the source S. S verifies the authenticity of the reply using the shared key between S and D ($K_{DS}$) and updates its routing table to the destination.

A → stores: RRP || $MAC_{K_{SD}}$(RRP) || $MAC_{K_{auth(D)}}$(RRP|| $MAC_{K_{SD}}$(RRP)).

D → broadcasts: $K_{auth(D)}$.

A → verifies the request-reply pair (Section 3.4.1).

A → updates its routing table to S an D.

A → C: RRP || $MAC_{K_{SD}}$(RRP) || $ID_A$ || $ID_C$ || $ID_D$ || $MAC_{K_{auth(A)}}$(RRP|| $MAC_{K_{SR}}$(RRP) || $ID_A$ || $ID_C$|| $ID_D$).

## 3.4 Intermediate Node Verification

Since the route discovery is achieved by flooding, a mechanism is required to limit the amount of flooding through the network. Each node only needs to broadcast the same request only once and it must suppress any further request copies. This can be achieved by attaching a unique identifier in the header of each new route request from a certain node. The identifier is also required to distinguish new requests from old replayed ones. *SRPS* uses a novel idea for the identifier to achieve intermediate node authentication of the route request and the route reply through a practical light weight protocol that prevents a malicious node from including itself in an *already established routes*.

### 3.4.1 Request-Reply Verification

Let SN be an increasing unique sequence number that is incremented with every new route request issued by a node. Let S be the route request source, D be the route request destination, and X be an intermediate node between S and D, which hears the route request from the intermediate node Y. Let N be the length of the hash sequence, and F represents the hash function.

#### 3.4.1.1 The First Route Request

*The route request initiator (A):*

1. Computes $v_0 = E_{K_{SD}}[SN]$ as the seed for the hash sequence
2. Computes the hash sequence $\{v_0, v_1, v_2, ..., v_n\}$ where $v_i = F(v_{i-1})$.
3. Broadcasts the route request holding SN and SNV = "$v_n \| n$."

*An intermediate node (B):*

1. Stores: $SN, v_n$ in addition to the information explained in Section 3.3.

*The route request destination (D):*

5. Computes $v_0 = E_{K_{SD}}[SN]$ as the seed for the hash sequence
6. Computes the hash sequence $\{v_0, v_1, v_2, ..., v_n\}$ where $v_i = F(v_{i-1})$.
7. Sends back to S holding the same SN and SNV = $v_{n-1}$

*An intermediate node (B):*

4. Verifies that $v_n = F(v_{n-1})$, and replaces $v_n$ with $v_{n-1}$
5. Sets the route table to S and D

So if the MAC values associated with the route request and the route reply and the hash values are verified to be correct, each intermediate node stores $v_{n-1}$ as a commitment for future route requests between S and D. The next RDP from S will carry $v_{n-2}$ and the corresponding RRP will carry $v_{n-3}$. A malicious node will not be able to compute $v_{n-2}$ or $v_{n-3}$ thus it will not be able to convince any of the nodes in the route to include itself in the route. In general the $i^{th}$ route request will follow the following protocol.

*3.4.1.2 The $i^{th}$ Route Request*

*The route request initiator (A):*

1. Broadcasts a route request holding the current sequence number SN and a SNV=$v_{n-2(i-1)} \| n-2(i-1)$

*An intermediate node (B):*

2. Stores: $SN, v_{n-2(i-1)}$ in addition to the information explained in Section 3.3.

*The route request destination (D):*

3. Sends back to S a route reply holding SN and $SNV = v_{n-2(i-1)-1}$.

*An intermediate node (B):*

4. Verifies that $v_{n-2i-1} = F(v_{n-2(i-1)-1})$ and replaces $v_{n-2i-1}$ with $v_{n-2(i-1)-1}$

5. Sets the route table to S and D

However, a DoS attack could be launched by colluding malicious nodes, to prevent S and D from establishing routes between them. A simple example of this attack is given in Figure 2. Let $M_1$ and $M_2$ be two malicious nodes, and S and D have not yet initialized a route path between them. $M_1$ could impersonate S and broadcasts an RDP packet claiming that it is from S. On the other end $M_2$ impersonates D and replied with a legitimate RRP. When S and D tries to establish a route between them, any node which is part of the fake route (the route between $M_1$ and $M_2$) will drop the request because it will not be authentic.

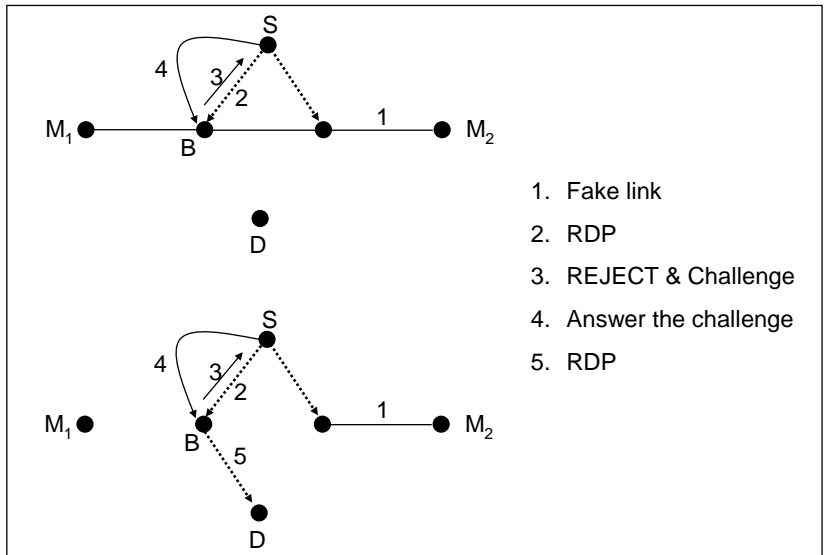

**Figure 2: DoS avoidance and Detection launched by impersonation**

1. Fake link
2. RDP
3. REJECT & Challenge
4. Answer the challenge
5. RDP

Three different solutions to prevent this DoS attack. The first method is through prevention by not allowing ID spoofing, so no malicious node will be able to impersonate another node; ID spoofing prevention is explained in Section 4. The second method is by requiring the source of the initial route request to be authenticated individually by each node in the route. This can be achieved by modifying the *Initial Route Request* protocol as follow:

### *3.4.2 First Route Request Version II*

*The route request initiator (A):*

1. Computes $v_0 = E_{K_{SD}}[SN]$ as the seed for the hash sequence
2. Computes the hash sequence $\{v_0, v_1, v_2, ..., v_n\}$ where $v_i = F(v_{i-1})$.
3. Broadcasts the route request holding SN and SNV = "$v_n \| n$."

*An intermediate node (B):*

4. Stores: $SN, v_n$ in addition to the information explained in Section 3.3.

*The route request destination (D):*

5. Computes $v_0 = E_{K_{SD}}[SN]$ as the seed for the hash sequence
6. Computes the hash sequence $\{v_0, v_1, v_2, ..., v_n\}$ where $v_i = F(v_{i-1})$.
7. Sends back to S a route reply holding SN and $SNV = v_{n-1}$

*An intermediate node (B):*

8. Verifies that $v_n = F(v_{n-1})$ and replaces $v_n$ with $v_{n-1}$.

9. Generates a random number $r_B$

10. Sends to S with the route reply "$E_{k_{SB}}[B, S, r_B]$"

*The route request initiator (A):*

11. Sends back to B "$E_{k_{SB}}[B, S, r_B + 1]$"

*An intermediate node (B):*

12. Sets the path to S and D if the reply of S is valid.

The third method to countermeasure the DoS attack is on-demand detection and a voidance. Let B be an intermediate node that rejects the legitimate route request of S due to the earlier established fake route by $M_1$. When S gets the route request from S, it rejects the request and sends a challenge back to S as "$E_{k_{SB}}[B, S, r_B]$", S will reply back by "$E_{k_{SB}}[B, S, r_B + 1]$". Thus B will accept the request and broadcast it. At the same time B will send back to S, the previous two hops towards the source of the fake route and the next hop towards the fake destination. S uses these links to trace the initiator of the fake route and thus detect it.

A final issue is the renewal of the hash sequence when the current sequence runs out. The following protocol is used to connect the hash sequence $\{u_0, u_1, u_2, ..., u_n\}$ with the current sequence $\{v_0, v_1, v_2, ..., v_n\}$ assuming that $v_3$ is the last disclosed key:

### 3.4.3 Hash Sequence Renewal Protocol

Let S be the route request source, D the route request destination, and B an intermediate node in the path between S and D.

$S \rightarrow D: E_{v_2}[u_n]$

$B: \text{stores } E_{v_2}[u_n]$

$D \rightarrow S: u_n$

$B: \text{stores } u_n$

$D \rightarrow S: v_2$

$B: \text{verifies that } h[v_2] = v_3 \text{ and computes } E_{v_2}[u_n] \text{ and compares it with the stored value}$

$B: \text{uses } u_n \text{ as the new authentication value}$

## 3.5 Neighbor monitoring

Each sensor node performs a neighbor watch by placing itself in the promiscuous mode and observes the header of the packets sent out by its neighbors. Figure 3 illustrates the concept of neighbor monitoring in which node S sends a packet P destined to node D. The packet P reaches node X in the path from the previous hop B. X must forward the packet to A, which must send it to D. When X forwards the packet, any one of the nodes M, N, and A will got it if that node does not have a collision.

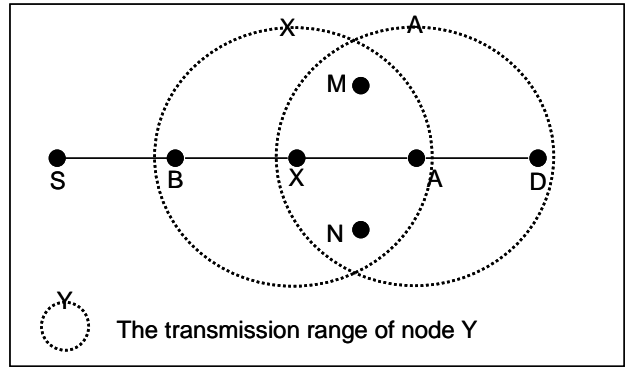

**Figure 3: An example of neighbor watch**

All the nodes that have both the sender and the receiver as neighbors are called the guards of the link between the sender and the receiver. In Figure 3, Nodes M, N, and X are guards for the link between X and A. Each guard of the link adds the packet information from the header of P in its watch buffer and time stamps it. The receiver, A, must forward the packet within a certain time threshold that depends on the density of the nodes, the bandwidth, and the traffic density. If a guard does not hear node A transmitting the packet within the time threshold, it will accuse A as performing a packet *drop*. If a guard hears node A transmitting the packet within the time threshold but detects a change in the packet content or header, it will accuse A as performing a packet *forge* or *change*. If a guard hears node A transmitting the packet, claiming that it got it from A, but they don't have the corresponding packet information, from the packet sent by X, in their watch buffer, they will accuse node A as performing packet *fabricating* or at least packet *delaying*.

## 3.6 Secure Disjoint Multi-path Discovery

Multipath protocols which look for maximally disjoint paths [49] are vulnerable to the tunneling attack which is a special case of the wormhole attack. In tunneling attack two malicious nodes collaborate to tunnel routing messages to one another so that a destination may falsely believes that two paths are disjoint while they share multiple nodes.

To help establish disjoint routes to the destination we use an idea inspired by Hu *et. al.* [37] which is used to prevent rushing attacks. In almost all the previously mentioned on demand ad-hoc and sensor network routing protocols, an intermediate node forwards the first announcement of a certain request and suppresses any following announcements. In *SRPS*, as we mentioned earlier, each node, say B, waits for a random a mount of time before forwarding the announcement that it heard. During that waiting time, it buffers all the announcements of the same request. At the same time, B listens to any neighbor, say E, whose timer times out and forwards one of the announcements he got, if that announcement comes from the same source as any one of those that B has in its buffer, then that announcement is suppressed from the buffer and is excluded from the random selection process done by B. Finally, when B's counter time-out, or when it gets a number of announcements greater than certain threshold value, it picks a random announcement from its buffer and forwards it.

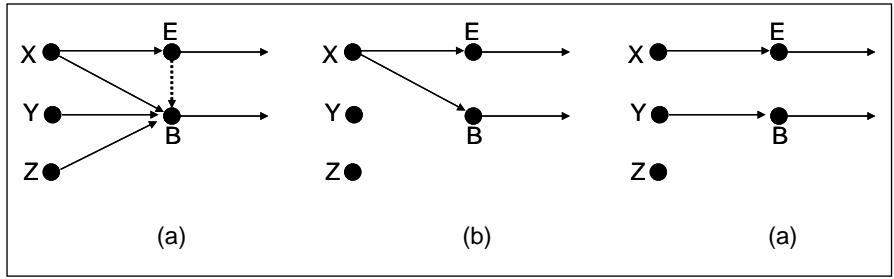

**Figure 4: (a) Route Request Collection at B and E; (b) E and B Forward the First Route Request they Get from X; (C) E and B Randomly Broadcast a Route Request.**

An example is shown in Figure 4 , let B receives route requests from nodes X, Y, and Z, and let E is a neighbor of B who also receives from X, and let the route request from X is the first to arrive to both B and E, Figure 4(a). If nodes B and E forward the first route request they got and drop the others they both will forward the route request they got from X as shown in Figure 4(b) which results in joint paths in node X. However, using the our technique, assuming that the timer of E runs out before that of B and that E broadcasts the message it received from X, then B will drop the X's packet. When B's timer runs out, it selects at random one of the Y, and Z packets and broadcasts it. The result paths are disjoint as shown in Figure 4(c).

When an intermediate node, say B, receives more than one reply for the same route discovery, which happens when two none-neighbor nodes forward the route request from the same previous hop. Consider for example the scenario shown in Figure 5. Node B had forwarded the route request it got from node A. Both of the none-neighbor nodes X and Y received and forwarded the route request they got from B.

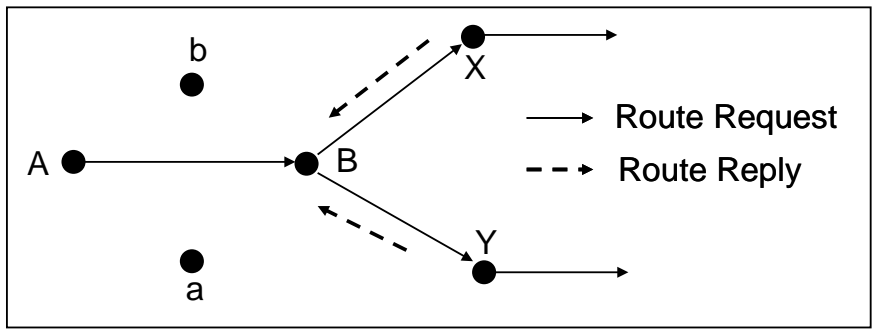

**Figure 5: Route Reply Behavior of an Intermediate Node**

As shown in Figure 5, B got a route reply back from both X and Y, assume that the X's route reply arrive first. Four different scenarios occur when B got the second route reply, Y's route reply, based on the behavior of B. (i) if B is an honest node, it will drop the second route reply, (ii) if B is malicious and A is an honest node, even if B forwards the second route reply to A, in attempt to include itself in both routes, A will drop it, (iii) if B is malicious and A is malicious, and B forwards the route reply to A, this moves us recursively to step (i) where node A takes the role of node B, and the next hop from A plays the role of node A, (iv) if B is malicious and it forwards the second route reply to another node, say $\alpha$, B will succeed in including it self in two "different routes". To countermeasure this attack, we use the neighbor watch. When a node, say $\beta$, overhears a neighbor forwarding a route reply it will save the route reply information in its watch buffer for a certain time $\tau$. If $\alpha$ overhears the same neighbor, B, forwarding the same route reply again within $\tau$, it will accuse that neighbor as trying to include himself in multiple disjoint routes, and thus delete any routing entry it may have between the source and destination that includes B. Thus in Figure 5, node A upon overhearing node B forwarding the next route reply will delete the first route reply from its routing table and thus preventing B from being in two routes. $\tau$ is selected to be the same as the threshold time after which the initiator of the route request will no longer accept any new route replies.

The initiator of the route discovery gives a priority level for each route reply it receives; the priority that we propose is based on the minimum delay. So the faster reply will be considered the highest priority one, even it may come through a relatively longer route (more number of hops). The rationale behind that is a longer route with less congestion or less malicious behavior is better than a shorter and congested route.

### 3.7 Route Maintenance

If an already established route between the source, S, and the destination, D, is broken either naturally; e.g. a node on the route exhausted its power, or maliciously by a compromised node in the path dropping data packets, then S has three options. The first is to revert to an alternate disjoint route if it has one. The second is to initiate a new route discovery process. The third is to ask the nodes at the edges of the faulty link to handle the problem and discover alternate routes to pass the traffic between them. For example if the route contains S-B- ~~~~ -X-Y-Z-~~~-R, and if Y is the faulty node, then S asks X to find an alternate route to Z and ask both X and Z to update their routing tables to S and D accordingly.

The ability of S to discover the broken path depends on the nature of the connection between S and D. If S is used to receive an acknowledgment through this route it will notice the problem by failing to receive the appropriate acknowledgments. If D expects to receive data from S and that data did not arrive,

D may request the data from S through a different route since it may guess the problem. If neighbor monitoring (Section 3.5) of data packet is enabled, the guards of the broken link will detect the failure and report it to S.

## 4 Security Analysis

In this section we will show how *SRPS* mitigates known attacks against the routing infrastructure.

Conjecture: *SRPS* route discovery does not allow any route to be established through a wormhole, Sybil, Sinkhole, rushing, or HELLO flood attacks.

*Conjecture#1*: *SRPS* does not allow any alteration, or replaying of route information

*Proof:* The route request and the route reply are both authenticated by the source and the destination of the route using a shared key known only to them. No malicious node can generate or change a valid route request that can be accepted by the destination, and no malicious node can generate or change a route reply that can be accepted by the initiator of the corresponding route request. The increasing sequence number associated with each route request-reply pair prevents replaying of old route requests and replies. The intermediate node verification prevents any number of colluding malicious nodes from including a malicious node in an already established route using the SN and the SNV as explained in Section 3.4.

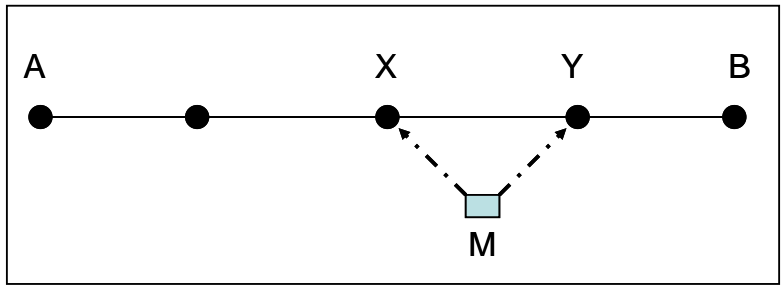

**Figure 6: An Example of a Malicious Node Trying to Include Itself in an Established Route.**

For example, assume a path is established between A and B as shown in Figure 6. Let M be a malicious node that tries to include itself in this path. For M to be able to do so it must convince X to change its routing table to point to M for packet destined to B, and convince A to change its routing table to point to M for packet destined to A. So M has to generate a valid route request or replay an old request, other wise the intermediate nodes will ignore the request since it will not be correctly verified. If M replays an old route request, the target B will discover that through the old sequence number that it holds and thus ignore the request so X and A will not receive a reply and thus will not update their routing tables. For M to generate a valid request it must know the key shared between A and B, which is provably infeasible unless either A or B or both are compromised, in that case there is no need even to try this attack. M as well may try to offer a faster service to A by claiming that it is one hop away from B, may be by using a high power transmission, but neither A nor B will believe it because it is not in the neighbor list of either of them.

*Conjecture#2*: *SRPS* does not allow any malicious node to establish a Sinkhole.

*Proof:* In a Sinkhole, an attacker tries to attract all the traffic from a particular area through a malicious node by including the malicious node in the route to all nodes in that area. If succeeded, the attacker can launch other attacks such as blackhole (blocking) or selective forwarding of data traffic. This attack typically works by making the malicious node look especially attractive for the surrounding nodes. For example an attacker could replay or spoof a high quality route to the destination. Some countermeasures of this attack rely on an end-to-end acknowledgments containing reliability or latency information. However, a powerful attacker who has enough power to transmit the packet directly to the target could easily defeat these countermeasures.

*SRPS* is not vulnerable to the sinkhole attack, due to the local knowledge of each node and due to intermediate node verification. A malicious node can't interact with any non-neighbor node; the malicious node communication will be rejected because it is not a neighbor node. This means that a malicious node can't shout loudly to convince far nodes that they are only one hop a way from it and thus attracting their traffic to it. Also a malicious node can't include it self in an established route or spoof or alter or replay any route information as proved in conjecture#1.

*Conjecture#3*: *SRPS* does not allow any malicious node to spoof routing information.

*Proof:* a malicious node, say M, has two possibilities to be able to spoof routing information; (i) generate the routing information and claim the ID of the victim node, say S. Any neighbor of M that does not have S as a neighbor will automatically reject the spoofed data. Any neighbor of M that is also a neighbor of S will also reject the data since it will not be verified correctly using the neighborhood authentication mechanism introduced in Section 3.2. (ii) Generate the routing information and claim that it has received it from S through another node, say B. If B is not a neighbor to M, then all the neighbors of M will automatically reject the spoofed data. If B is a neighbor to M, then the guards of the link between M and be will detect the forge since they do not have the corresponding data (from B to M) in their watching buffer.

*Conjecture#4*: *SRPS* does not allow any malicious node to launch a Sybil attack for routing purposes.

*Proof:* In Sybil attack, a malicious node presents multiple identities to the network [22]. This attack is especially destructive for protocols that are used to discover disjoint multiple routes; a malicious node can include itself in multiple different routes by presenting different identities to other nodes. As proved in Conjecture#3, *SRPS* prevents ID spoofing, so no malicious node can present an identity other than its own to the network and thus can't launch the Sybil attack.

*Conjecture#5*: *SRPS* does not allow any colluding malicious nodes to launch a Wormhole attack.

*Proof:* The wormhole attack [35],[48][35] involves two distant colluding malicious nodes to understate their distance from each other. It is more effective when used to create sinkholes or artificial links that attract traffic. This attack can be launched either by using an out-of-bound channel available only to the attacker or by tunneling messages received in one part of the network and replaying them in a different part. It is so effective that it can be launched even without having access to any cryptographic keys.

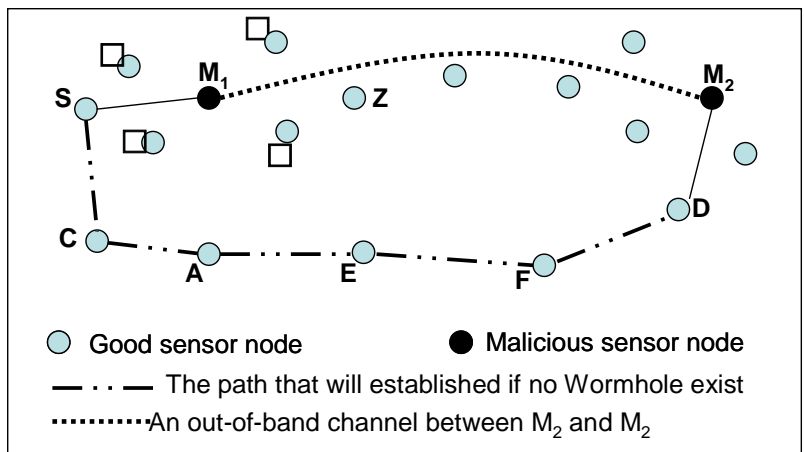

**Figure 7: Wormhole illustration example**

*SRPS* detects and defeats the wormhole attack; neighborhood authentication (Section 3.2) defeats the possibility of launching the attack by nodes that does not have cryptographic keys. So for this attack to work in *SRPS*, the two colluding nodes must be compromised nodes owning legitimate cryptographic keys. We consider the scenarios by which wormhole attack could be launched. Two colluding nodes use an out-of-band invisible channel to the underlying sensors or packet encapsulation to tunnel routing information between them, Figure 7. When $M_1$ hears the route request packet initiated by S, it directs it to $M_2$. $M_2$ rebroadcasts the packet to its neighbors and eventually it reaches the target D. D then generates a route reply and sends it back until it reaches $M_2$. $M_2$ sends the route reply back to $M_1$ using the unseen channel between them. $M_1$ forwards the route reply back to S and it must append to the header the ID of the previous hop from which it got the route reply. $M_1$ has two choices, either to say the truth and append the ID of $M_2$ as the previous hop or lie and append the ID of one of its neighbors, say Z, as the previous hop. In the first choice node S will reject the route reply because it knows that $M_2$ is not a neighbor to $M_1$, so $M_2$ can't be the previous hop from $M_1$, also all the neighbors of $M_1$ will detect the malicious activity of

$M_1$. In the second case, all the guards of the link between Z and $M_1$ (Z,α, and β) will detect B as forging the route reply since they don't have the corresponding information from Z in their watch buffer. So in both cases $M_1$ will fail to pass the route reply and thus the wormhole will not succeed.

*Conjecture#6*: *SRPS* prevents rushing attacks.

*Proof:* In the rushing attack, an adversary who hears the route request broadcast rushes to rebroadcast the request in attempt to make the route request broadcasted by him the first to reach all the neighbors of the destination. If the attacker succeeds in doing that, then any route discovered by this route discovery will include a hop through the attacker. As a result the attacker can easily launch a DOS attack and prevent the source from discovering any usable routes to the destination.

*SRPS* inherently implement and use rushing attack prevention (RAP) [37] for discovery of multiple disjoint routes, Section 3.6. An intermediate node will not forward the first route request it got (may be from a rushing malicious node), but it will collect a number of route requests from different neighbors and randomly select one of them to rebroadcast.

## 5 SRPS analysis

### 5.1 Coverage analysis

In this Section, we characterize the probability of miss detection and false detection as the network density increases and the detection confidence index, $\gamma$, varies. Results provide some interesting insight. For example, we are able to compute the required network density $d$ to detect $p$% of the attacks when the detection confidence index equals to $\gamma$.

Consider any two randomly selected neighbor nodes, $S$ and $D$, as shown in the Figure 8(a). $S$ and $D$ are separated by a distance $x$, and the communication range is $r$. The guard nodes for the communication between $S$ and $D$ are those nodes that lie within the communication range of $S$ and $D$, the shaded area in Figure 8(a). The area of the shaded region in Figure 8(a) is given by $Area(x) = 2r^2 \cos^{-1}\left(\dfrac{x}{2r}\right) - (2x)\sqrt{r^2 - \dfrac{x^2}{4}}$. The value of $x$ is from a uniform random variable which has the range (0, r). Thus the expected value of the area is given by $E[Area(x)] = \int_0^r \left\{ 2r^2 \cos^{-1}\left(\dfrac{x}{2r}\right) - (2x)\sqrt{r^2 - \dfrac{x^2}{4}} \right\} \dfrac{1}{r} dx = \sqrt{3} r^2$.

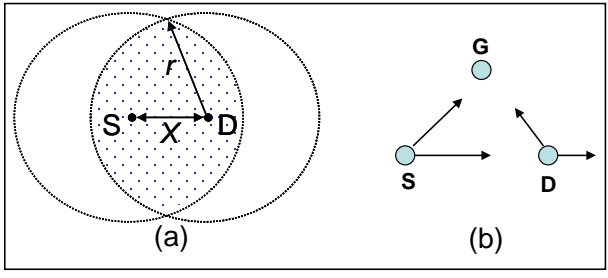

**Figure 8: (a) The area from which a node can guard the link between *S* and *D*; (b) Illustration of detection accuracy**

The minimum value of the Area(x), Areamin, is when x = r. The minimum number of guards, gmin, which can watch the link between S and D is given by $g_{min} = Area_{min} d = 0.36 r^2 d$, where $Area_{min} = Area(r)$. The expected number of guards, g, which can watch the link between S and D is given by $g = E[Area(x)]d = (\sqrt{3} r^2) d = \sqrt{3} r^2 d$. The number of neighbors of a node is given by $NB = \pi r^2 d$, thus, $g = \frac{\sqrt{3}}{\pi} NB \approx 0.55 NB$.

Based on the performance analysis of the IEEE 802.11 MAC protocol [51], we assume that each packet collides with constant and independent probability, $P_C$, i.e. $P_C$ is the probability of a collision seen by a packet being transmitted on the channel. Thus, each guard receives a packet with probability $\alpha = 1 - P_C$. Assume that $\mu$ malicious activities occur within a certain time window, *T*. Assume that a guard must detect $\beta$ malicious activities to cause the Mal$_C$ for a node to cross the threshold, and thus, generates an alert. Then, the alert probability at a guard is given by $P_{\beta|\mu} = \sum_{i=\beta}^{\mu} \binom{\mu}{i} \alpha^i (1-\alpha)^{\mu-i}$. The probability $p_\gamma$ that $\gamma$ of the guards detect the malicious node is given by $p_\gamma = \binom{g}{\gamma}(P_{\beta|\mu})^\gamma (1 - P_{\beta|\mu})^{g-\gamma}$.

The probability that at least $\gamma$ of the guards generate an alert, and thus detect the malicious node is given by

$$p_{\geq \gamma} = \sum_{i=\gamma}^{g} \binom{g}{i} (P_{\beta|\mu})^i (1-P_{\beta|\mu})^{g-i} = \frac{B(P_{\beta|\mu}, \gamma, g-\gamma+1)}{B(\gamma, g-\gamma+1)} = \frac{g!}{(\gamma-1)!(g-\gamma)!} \int_0^{P_{\beta|\mu}} u^{\gamma-1}(1-u)^{g-\gamma} du$$

Where $B(\gamma, g-\gamma+1)$ is the Beta function and $B(P_{\beta|\mu}; \gamma, g-\gamma+1)$ is the incomplete Beta function.

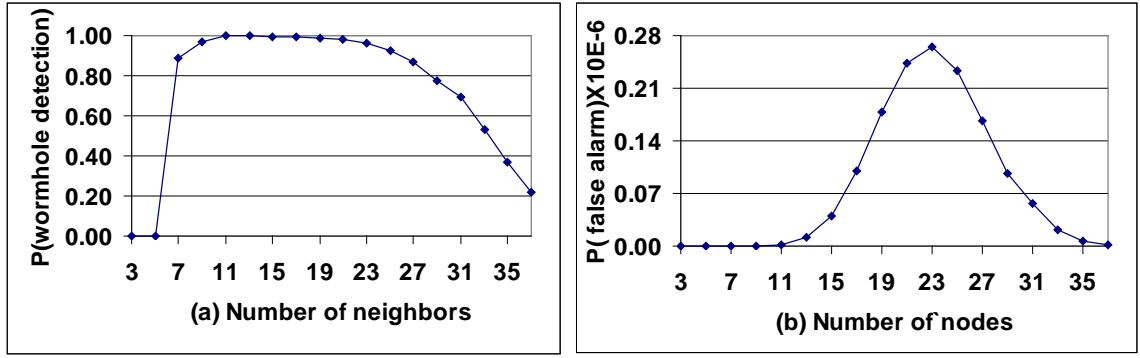

**Figure 9: (a) Probability of wormhole detection as a function of the number of neighbors; (b) Probability of false alarm as a function of the number of neighbors**

Figure 9(a) shows the probability of detecting the wormhole using $\mu = 7$, $\beta=5$, $\gamma=3$, and $P_C = 0.05$ when the number of neighbors is 3, and the number of compromised nodes $M = 2$. $P_C$ is assumed to increase linearly with the number of neighbors. Since the number of guards increases as the number of neighbors increases, the probability of detection increases since it becomes easier to get the degree of confidence required. However, the collision probability also increases as the number of neighbors increases, and thus the probability of detection starts to fall rapidly beyond a point. Figure 12, shows for the same $\mu$, $\beta$, and $P_C$ the probability of wormhole detection as a function of $\gamma$ when $NB = 15$ and $M = 2$. As the detection confidence index increases, the probability of detection decreases.

As shown in Figure 8(b), a guard $G$ will not detect a fabricated packet sent by $D$, claiming it was received from $S$, if $G$ experienced a collision at the time when $D$ transmits, thus, the probability of misdetection is given by $P_{MD} = \alpha = 1 - P_C$. A false alarm occurs when $D$ receives a packet sent from $S$, while $G$ does not receive that packet, and later, $G$ receives the corresponding packet forwarded by $D$. Thus, the probability of false alarm, $P_{FA}$, is given by $P_{FA} = P_C(1-P_C)^2 = (1-\alpha)\alpha^2$. Assume that S sends to $D$ $\mu$ packets, to be forwarded by $D$, within a certain time window, $T$, then the probability that $\beta$ or more of them are falsely detected is given by $P_{FA(\beta|\mu)} = \sum_{i=\beta}^{\mu} \binom{\mu}{i}(P_{FA})^i (1-P_{FA})^{\mu-i}$, and the probability that $\gamma$ or more guards send false alarms, leading to the node being flagged malicious, is given by

$$p_{FA \geq \gamma} = \sum_{i=\gamma}^{g}\binom{g}{i}(P_{FA(\beta|\mu)})^i(1-P_{FA(\beta|\mu)})^{g-i} = \frac{\beta(P_{FA(\beta|\mu)}, \gamma, g-\gamma+1)}{\beta(\gamma, g-\gamma+1)} = \frac{g!}{(\gamma-1)!(g-\gamma)!} \int_0^{P_{FA(\beta|\mu)}} u^{\gamma-1}(1-u)^{g-\gamma} du$$

Figure 9(b) shows the probability of false alarm as a function of the number of nodes for the same parameters as in Figure 9(a). The non monotonic nature of the plot can be explained as follows. As the number of neighbors increases, so does the number of guards. Initially, this increases the probability that at least $\gamma$ will miss the packet from S to the guard but not from D to the guard, leading to false detection at these γ guards. But beyond a point, the number of neighbors causes increased contention leading to the probability that both these packets will be missed at the guard and will thus not lead to false detection. The worst case false alarm probability is negligible (less than $0.3 \times 10^{-6}$).

## 5.2  Cost Analysis

In this section we analyze the resource requirements (memory, computation, and communication) of SRPS. To compare and judge these requirements we mention the resources available to one of the most common sensor nodes, the MICA motes. This mote has Atmega128 4 MHZ processor, 4 K byte RAM, 128 K byte flash memory (program memory), 512 K byte nonvolatile memory and 38Kbps bandwidth.

### 5.2.1  Memory Cost

Each node needs to store the list of neighbors, the neighbors of each neighbor, a commitment key for each neighbor, a commitment string for itself, a routing table, and a watch buffer. The number of neighbors, $Nn$, depends on the density of the network.. The length of the commitment string, $Lc$, depends on the available storage and it can be as small as two elements and as large as desired. The more elements in the string, the less communication overhead required for commitment renewal. The size of the routing table depends on the number of routing table entries, $RTE$, and the size of each entry. Each routing table entry consists of the final target identity, the neighbor identity that leads to the destination, and the sequence number of the route request that builds the route entry. The watch buffer size depends on the number of buffer entries, $NBE$, and the size of each entry. Each watch buffer entry consists of 4-node identities (one for the source, one for the target, and 2 for the previous two hops), an $SN$, and an $SNV$. For example, let the identity size = 4 bytes, the MAC size = 10 bytes, the key size = 8 bytes, the $SN$ = 4 bytes, $SNV$ = 10 bytes, then the required storage will be $Nn (4 + 8) = 12Nn$ bytes for the neighbor list, 64 $Lc$ for the commitment string, $RTE (4+4+4) = 12RTE$, and $NBE(4+4+4+4+4+10) = 30NBE$. The total amount of memory required = $12\ Nn + 64\ Lc + 12\ RTE + 30\ NBE$ bytes. If $Nn = 20$, $Lc = 10$, $RTE = 20$, and $NBE = 10$, then the memory requirement will be 1420 byte which is less than 1.5 kilobytes.

### 5.2.2  Computational Cost

The initiator needs to calculate three MAC values; one for end-to-end verification for the target to verify the authenticity of the source using the shared key between the source and the destination, another is used for neighborhood authentication using the current commitment key, the other is used calculate the $RSN$ using the shared key between the source and the destination over the current sequence number. The initiator also calculates a HASH value to find $SNV$. And another HASH value to generate the next commitment key. Each intermediate node needs to verify the commitment key (one HASH) calculation, verify the neighborhood authenticity of the packet (one MAC calculation), calculate the next commitment key (one hash calculation), authenticate the packet (one MAC), and if the intermediate node is in the path of the route reply then that intermediate node needs to verify the request-reply pair (one HASH). The destination needs to verify the source (one MAC), the neighbor hood (one MAC), calculate the RSN (one MAC), sign the reply (one MAC), calculate the next commitment key (one HASH) sign the packet for its neighbors verification (one MAC). So the source needs to calculate 3 MAC values and 2 hash values. The intermediate node needs to calculate 2 MAC values and 3 hash values. And the destination needs to calculate 5 MAC and 1 HASH values. The following table provides the overhead required by some cryptographic algorithms [47].

|  | **RC4** | **RC5** | | **AES** |
|---|---|---|---|---|
|  |  | 5 Rounds | 12 Rounds |  |
| Speed (128bits/ms) | 1.299 | 5.471 | 12.475 | 102.483 |
| Data size (Bytes) | 258 | 68 | 124 | 1165 |
| Code size (bytes) | 580 | 1436 | 1436 | 9492 |

**Table 1: Cryptographic algorithms costs**

### 5.2.3 Communication Cost

The source node needs to broadcast the RDP and the key commitment packet. Each intermediate node needs to transmit the same amount of traffic as the source node. The destination needs to transmit the RRP and the commitment key, and each intermediate node in the route to the source needs to transmit the same amount of traffic. The size of the RDP packet = 4 (source ID) + 4 (destination ID) + 4 (previous hop ID) + 1 (flag) + 4 (*SN*) + 10 *SVN* + 10 (end-to-end MAC) + 10 (neighbor MAC) = 47 bytes. The commitment key packet size = 4 (sender ID) + 8 (the key) = 12 bytes. The RRP packet size = 4 (source ID) + 4 (destination ID) + 10 (*RSN* =18 bytes).

# 6 Simulation Results

We use the ns-2 simulation environment to simulate a data exchange protocol, individually in the baseline case without any protection, and also with SRPS. We distribute the nodes randomly over a square sensor field with a fixed average node density. Thus, the sensor field size varies (80×80 m to 204×204 m) with the number of nodes. We assume that the route is evicted from the cache after a timeout period expires ($TOut_{Route}$). We simulate the wormhole attack and study its consequences on the network with SRPS and without SRPS. When a malicious node hears a route request, it directs the request to all the other malicious nodes in the network using an out-of-band channel or using packet encapsulation. For packet encapsulation, we assume that the colluding nodes always have a route between them. We simulate the Wormhole attack. We simulate the out-of-band channel by letting the compromised nodes deliver the packets instantaneously to their colluding parties. The wormhole attack exercises the principal features of SRPS, namely, local monitoring and are more difficult to mitigate than other attacks. Hence, we simulate it in preference to other attacks. After a wormhole is established, the nodes drop any data packet going over that wormhole.

Each node acts as a data source and generates data using an exponential random distribution with inter-arrival rate of $\mu$. The destination is chosen at random and is changed using an exponential random distribution with rate $\xi$. The important input parameters to the simulation are the detection confidence index ($\gamma$), the number of neighbors for each node (*NB*), which is a function of the node density, the number of nodes in the network (*N*), and the number of compromised nodes (*M*). The output parameters include the isolation latency, the number of data packets generated, the number of data packet dropped due to the wormhole, the number of routes established, and the number of routes affected by the wormhole. The simulation also accounts for losses due to natural collisions. The isolation latency is calculated from the time a malicious node starts a wormhole attack until it is completely isolated by all of it neighbors. The guards inform all the neighbors of the detected malicious node through multiple unicasts. The output parameters that we present here are obtained by averaging over 30 runs. For each run, the malicious nodes are chosen at random such that they are more than 2 hops away from each other.

Table 2 summarizes the range of input parameter values for the experiments conducted.

| Parameter | Value | Parameter | Value | Parameter | Value |
| --- | --- | --- | --- | --- | --- |
| Tx Range (r) | 30 m | $\gamma$ | 2-8 | N | 20,50,100,150 |
| NB | 8 | $\mu$ | 1/10 sec | $\xi$ | 1/200 sec |
| $TOut_{Route}$ | 50 sec | M | 0-4 | Channel bw | 40 kbps |
| $\beta$ | 5 | $\tau$ | 0.5 sec | T | 200 |

**Table 2: Input parameter values for SRPS simulations**

Figure 10 shows the number of packets dropped as a function of simulation time for the 100-node setup with 2 and 4 colluding nodes both with SRPS and without SRPS. *Since the number is vastly different in the two cases, they are shown on separate Y-axes, the axis on the left of each figure corresponds to the baseline case (without SRPS) and the axis to the right corresponds to the system using SRPS.* In the baseline case, since wormholes are not detected and isolated, the cumulative number of packets dropped continues to increase steadily with time. But in the SRPS case, as wormholes are identified and isolated for good, the cumulative number stabilizes. Notice that the cumulative number of packets dropped grows for some time even after the wormhole is locally isolated, due to the cached routes that contain the wormhole and continue to be used till route timeout occurs.

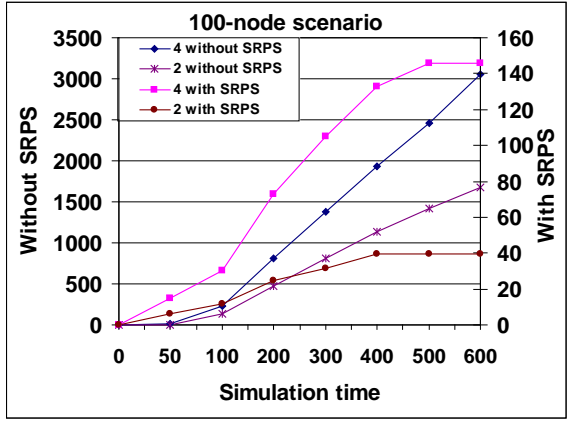

**Figure 10: The cumulative number of dropped packets with and without SRPS**

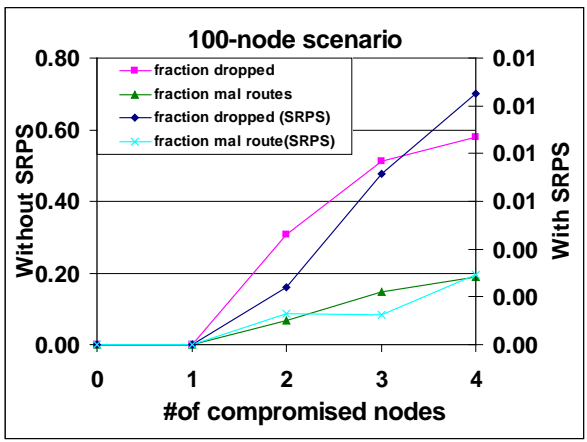

**Figure 11: Fraction of dropped packets and malicious routes with and without SRPS**

Figure 11 shows a snapshot, at the simulation time of 2000 secs, of the fraction of the total number of packets dropped to the total number of packets sent and the fraction of the total number of routes that involve the wormhole to the total number of routes established. This is shown for for 0-4 compromised nodes for both scenarios — with SRPS and without SRPS. With 0 or 1 compromised node, there is no effect on normal traffic since no wormhole is created. Notice that the relationship between the number of dropped packets and the number of malicious routes is not linear. This is because the route established through the wormhole is more heavily used by data sources due to the aggressive nature of the malicious node at the end of the wormhole. If we track these same output parameters over time, with SRPS, they would tend to zero as no more malicious routes are established or packets dropped, while without SRPS they would reach a steady state as a fixed percentage of traffic continues to be affected by the undetected wormholes.

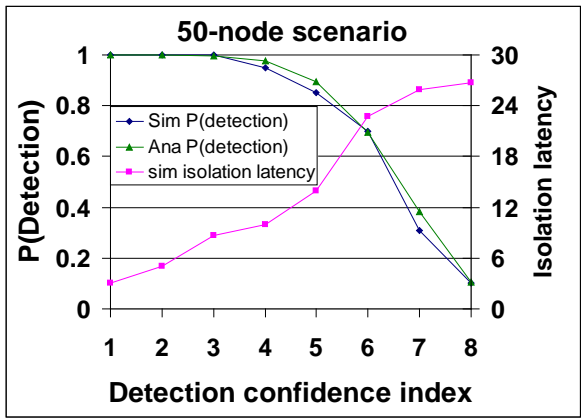

**Figure 12: Detection probability and latency with variations in detection confidence index for SRPS**

Figure 12 bears out the analytical result (Figure 9) for the detection probability as the detection confidence index ($\gamma$) is varied with NB = 15 and M = 2. We also show the isolation latency. As $\gamma$ increases, the detection probability goes down due to the need for alarm reporting by a larger number of guards, in the presence of collisions. Also the isolation latency goes up, though it is very small (less than 30 s) even at the right end of the plot.

## 7  Conclusion

We have presented a secure routing protocol, called SRPS, for resource constrained sensor networks. SRPS represents a stand alone lightweight protocol that mitigates all the known attacks namely, traffic blocking, HELLO flood, sinkhole, wormhole, Sybil, rushing, spoofed, altered, and replayed attacks. We present a detailed security analysis of SRPS and show its ability of dealing with each one of these attacks. SRPS exploits novel ideas to achieve, in addition to the end-to-end authentication, an intermediate node verification using the specially generated sequence numbers and sequence number verifications. Neighbor watch and wait-while-collect are used to help in defeating these attacks and establishing multiple disjoint routes.

We present mathematical analysis of the detection coverage in SRPS, and the resource overhead that SRPS requires and show that these requirements are within the available limits for the current sensor technology. Finally we provide simulation results that show the capabilities of SRPS in mitigating the wormhole attack.